\documentclass{cimento}
\usepackage{graphicx}
\usepackage[latin1]{inputenc}

\title{Equilibrium configurations of relativistic White Dwarfs}
\author{G.~Bertone \atque R.~Ruffini }
\instlist{\inst ICRA, International Centre for Relativistic Astrophysics, 
Dipartimento di Fisica,
Universit\`{a}  di Roma I-00185 Roma Italy
}

\PACSes{\PACSit{00.00}{ }
\PACSit{---.---}{\ldots}}
\begin{document}

\maketitle

\begin{abstract}
The Feynman-Metropolis-Teller treatment for compressed atoms \cite{feymete} is here reconsidered in the framework of the relativistic generalised Fermi-Thomas model, obtained by Ruffini \textit{et al.} \cite{ruffini}. Physical properties of a zero temperature plasma is thus investigated and the resulting equation of state, which keeps into account quantum, relativistic and electromagnetic effects, is applied to the study of equilibrium configurations of relativistic White Dwarfs. 

It is shown that numerical evaluation of such configuration leads, for the same central density $\rho_c$, to smaller values of radius R and of mass M than in the classical works of Chandrasekhar \cite{chandra} and Salpeter \cite{salpeter2}, the deviations being most marked at the lowest densities (up to 30\% from the Chandrasekhar model and 10\% from the Salpeter one for $\rho_c \sim 10^6g/cm^3$, corresponding to $M \sim 0.2 M_{\odot}$). 

At high densities we considered the occurrence of inverse beta decays, whose effect is to introduce gravitational instability of the configurations. We consequently find the maximum mass of White Dwarfs, which, for an Oxygen and an Iron WD, is respectively $1.365   M_{\odot}$ and $1.063   M_{\odot}$.
\end{abstract}

\section{Introduction}

That white dwarfs cannot have an arbitrary large mass and that a critical mass against gravitational collapse must exist in their configuration of equilibrium reachable by increasing their central densities was clearly understood in the thirties independently by Chandrasekhar\cite{CH1,CH2} and by Landau.\cite{L} 

Actually an  interesting alternative approach, with respect to the evaluation of the interaction term between the electrons and the nuclei, was advanced two years earlier than the work of Chandrasekhar and Landau  by the Soviet physicist Yacov Ilich Frenkel.\cite{FRK} This paper, quoted by Chandrasekhar in his classic book,\cite{chandra} has been in the past and is still today surprisingly ignored. Frenkel proposed to use a relativistic Thomas-Fermi model within a Wigner-Seitz approximation in order to describe stellar matter. The astrophysical motivations were not  clear to Frenkel, though the theoretical formulation he proposed would have deserved a much more thorough examination.

An important turning point came in this field due to the fundamental work of Feynman, Metropolis and Teller. While at los Alamos, they considered the equation of state of compressed matter\cite{feymete}.

It was Ed Salpeter\cite{salpeter,salpeter2}  who first applied the Frenkel equations and the Feynman-Metropolis-Teller\cite{feymete} approach to the treatment of white dwarfs:  Salpeter's  treatment was somewhat affected by the intricacies of the numerical integrations and the drastic approximations he adopted in the numerical solutions.

Today with relatively simple numerical computations, we can reconsider the problem by using a relativistic generalisation of gravitational Fermi-Thomas like equations \cite{ferre,ruffini,mrt} and present a comprehensive treatment by first considering the issue of the electromagnetic interactions between nuclei and electrons and turning later to the issue of inverse beta decays. The treatment presented here leads to the disappearance of all scaling laws with the chemical composition of the stars, which have been commonly assumed in all treatments up to now. A  new value of the critical mass of white dwarfs is obtained, smaller than the one originally proposed by Chandrasekhar and by Salpeter, and one which strongly depends on the chemical composition of the star. These results appear in principle to be observationally testable. Particularly attractive is the possibility of explaining the observed masses of binary pulsars: these systems are clearly neutron stars which could be formed by the onset of instability of stellar cores very close to the critical mass of white dwarfs as presented by Chandrasekhar\cite{giacco}. The evaluation of their masses will be clearly affected by our computations. 

Here, following Ferrerinho \textit{et al.} \cite{ferre} and Ruffini \textit{et al.} \cite{ruffini} we reconsider the Feynman-Metropolis-Teller\cite{feymete} approach in the framework of a generalisation of the Fermi-Thomas model, which is here briefly derived and solved (section 2) for different atomic species in different states of compression. Results so obtained are applied to determine the equation of state (section 3) of cold degenerate stars (White Dwarfs) to show how the relativistic and electromagnetic effects acting on microscopic scale affects the overall structure of the star. Finally, the equation of state obtained is used to determine (section 4) the equilibrium configurations of White Dwarfs and in particular the maximum mass configurations for different chemical compositions. A comparison is made with the classical works of Chandrasekhar and Salpeter to show the differences introduced by the use of the new equation of state, and to show how that they correspond respectively to a zero-order and first-order approximation of the complete treatment here presented.

\section{The generalised Fermi-Thomas model}

Let us consider the spherically symmetric problem of a nucleus with Z protons and A nucleons interacting with a fully degenerate gas of Z electrons. 

The fundamental equation of electrostatics for this problem reads, with usual meaning of symbols, 
\begin{equation}
\Delta V(r)=4 \pi e n_e(r)-4 \pi e n_p(r)
\label{elettro}
\end{equation}
where we introduced the number density of electrons $n_e(r)$ and of protons $n_p(r)$.

In the equation \ref{elettro} the quantities V(r) and n(r) are of course not independent, being related by  the equilibrium condition for a relativistic gas in a coulomb external potential (see ~\cite{lali})
\begin{equation}
c\sqrt{p_{F}^{2}+m^2c^2}-eV(r) = \mbox{const} \equiv E_F
\label{equili}
\end{equation}
where the name $E_F$ stands for \emph{Fermi-Thomas chemical potential} or \emph{Fermi Energy} of the electrons. 

To put the equation \ref{elettro} in a simple and adimensional form we introduce the new function $\Phi(r)$, related to the coulomb potential by the formula
\begin{equation} 
\Phi(r)=V(r)+E_F/e
\label{4}
\end{equation}
and the corresponding adimensional function $\chi$, implicitly defined by 
\begin{equation}
\Phi(r)=\frac{Ze\chi}{r}
\label{6}
\end{equation}

Furthermore we introduce the new independent variable x, obtained by the radius r with the relation  $r=bx$, where we put
\begin{equation}
b=(3\pi)^{2/3}\frac{\hbar^2}{me^2}\frac{1}{2^{7/3}}\frac{1}{Z^{1/3}} 
\label{eq:6}
\end{equation}

Using \ref{4} it is possible to put eq. \ref{equili} in the form
\begin{equation}
p^2_F=\frac{e²}{b} \Phi²+ 2me \Phi
\end{equation}
which, using \ref{6},becomes 
\begin{equation}
p_F=2mc\left(\frac{Z}{Z_{cr}}\right)^{2/3}\left(\frac{\chi}{x}\right)^{1/2}\left[1+ \left(\frac{Z}{Z_{cr}}\right)^{4/3}\frac{\chi}{x} \right]^{1/2} 
\label{impu}
\end{equation}
where
\begin{equation}
Z_{cr}=\left(\frac{3\pi}{4}\right)^{1/2}\left(\frac{\hbar c}{e^2}\right)^{3/2}\approx 2462.4 
\end{equation}

\begin{figure}
\begin{center}
\begin{tabular}[t]{|c|c|c||c|c|c|}

\hline 
\hline

$x_0^{He}$    &  $\chi^{He}(x_0)$  &   $t^{He}(x_0)$ & $x_0^{Fe}$    &  $\chi^{Fe}(x_0)$  &   $t^{Fe}(x_0)$    \\ 
\hline
&\\
$8.736\cdot 10^{-2}$ & 23.034 & 	0.286  &$1.516\cdot 10^{-1}$ & 11.335 &	0.902 \\

$5.009\cdot 10^{-2}$ & 38.896 & 	0.499  &$1.309\cdot 10^{-1}$ & 12.618 &	1.046 \\

$2.465\cdot 10^{-2}$ & 69.214 & 	1.016  &$1.001\cdot 10^{-1}$ & 15.023 &	1.369 \\

$1.021\cdot 10^{-2}$ & 111.192 & 	2.455  &$5.516\cdot 10^{-2}$ & 20.021 &	2.488 \\

$8.750\cdot 10^{-3}$ & 117.410 & 	2.864  &$2.727\cdot 10^{-2}$ & 24.336 &	5.036 \\

$7.672\cdot 10^{-3}$ & 122.312 & 	3.267  &$2.104\cdot 10^{-2}$ & 	25.450 &	6.527 \\

$3.495\cdot 10^{-3}$ & 143.907 & 	7.172 &$1.615\cdot 10^{-2}$ & 	26.366 &	8.503 \\

$2.876\cdot 10^{-3}$ & 147.474 & 	8.714  &$1.376\cdot 10^{-2}$ & 	26.827 &	9.980 \\

$2.126\cdot 10^{-3}$ & 151.932 & 	11.789  &$1.105\cdot 10^{-2}$ & 	27.362 &	12.437 \\

$9.246\cdot 10^{-4}$ & 159.380 & 	27.110  &$9.226\cdot 10^{-3}$ & 	27.727 &	14.890 \\

$5.910\cdot 10^{-4}$ & 161.516 & 	42.414  &$6.942\cdot 10^{-3}$ & 	28.192 &	19.790 \\

$4.343\cdot 10^{-4}$ & 162.531 & 	57.714  &$6.178\cdot 10^{-3}$ & 	28.350 &	22.239 \\

$2.108\cdot 10^{-4}$ & 164.001 & 	118.906  &$5.566\cdot 10^{-3}$ & 	28.477 &	24.686 \\

$1.677\cdot 10^{-4}$ & 164.294 & 	149.500  &$4.645\cdot 10^{-3}$ & 	28.669 &	29.580 \\

$1.039\cdot 10^{-4}$ & 164.758 & 	241.278  &$3.986\cdot 10^{-3}$ & 	28.808 &	34.472 \\

$9.225\cdot 10^{-5}$ & 164.854 & 271.869  &$3.491\cdot 10^{-3}$ & 	28.913 &	39.363 \\

$8.293\cdot 10^{-5}$ & 164.938 & 302.461  &$2.796\cdot 10^{-3}$ & 	29.062 &	49.143 \\

$7.532\cdot 10^{-5}$ & 165.012 & 333.052  &$2.154\cdot 10^{-3}$ & 	29.202 &	63.807 \\

$6.900\cdot 10^{-5}$ & 165.080 & 363.643  &$1.868\cdot 10^{-3}$ & 	29.265 &	73.580 \\

   &   &  \\
\hline
\hline
\end{tabular}
\end{center} 
\caption{Results of numerical integration of generalised Fermi-Thomas equation for Helium (3 columns on left) and Iron. For each atomic specie we give the values of the adimensional function $\chi$ for different atomic sizes (i.e. for different $x_0$and the corresponding adimensional Fermi momentum.}
\end{figure}

So we obtained an expression for the Fermi momentum in terms of adimensional quantities, and if we remember the relation between the Fermi momentum and the number density of a degenerate fermion gas
\begin{equation}
n_e=\frac{p_{F}^{3}}{3\pi^2\hbar^3}
\label{2}
\end{equation}
we obtain the following expression
\begin{equation}
n_e=\frac{Z}{4\pi b^3}\left(\frac{\chi}{x}\right)^{3/2}\left[1+ \left(\frac{Z}{Z_{cr}}\right)^{4/3}\frac{\chi}{x} \right]^{3/2} 
\label{adens}
\end{equation}

Let us try to express the second term of the right-hand side of eq.\ref{elettro} in terms of adimensional quantities: we shall assume an homogeneous spherical nucleus, with a radius given by the approximate formula 
\begin{equation}
r_{nuc}=1.2A^{1/3} \: fm
\end{equation}

The number density of protons is therefore
\begin{equation}
n_{p}=\frac{3Ze}{4\pi r_{nuc}^3}\Theta\left(x_{nuc}-x \right)
\end{equation}
 
Finally we can write eq.\ref{elettro} in the form
\begin{equation}
\frac{d^2 \chi}{dx^2}=\frac{\chi^{3/2}}{x^{1/2}} \left[1+ \left(\frac{Z}{Z_{cr}}\right)^{4/3}\frac{\chi}{x} \right]^{3/2}-\frac{3x}{{x_{nuc}}^3}\Theta\left(x_{nuc}-x \right) 
\label{thomfer}
\end{equation}
where $x_{nuc}$ is the adimensional size of the nucleus ($r_{nuc}=bx_{nuc}$).

Equation \ref{thomfer} is what we call \lq \emph{Generalised adimensional Fermi-Thomas equation\footnote{Classical Fermi-Thomas model is easily recovered in the limit $\left(\frac{Z}{Z_{cr}}\right)^{4/3}\frac{\chi}{x} \rightarrow 0$ and $x_{nuc}\rightarrow 0$, i.e. in the limit of non-relativistic expression for the momentum and of point-like nucleus.}}\rq.

The first initial condition for this equation follows from the fact that $ \chi \propto r \Phi $ and therefore $\chi   \stackrel{r{\rightarrow}0}{\longrightarrow}0$, and so
\begin{equation}
\chi(0)=0 
\label{con1}
\end{equation}

The second condition comes from the normalisation condition 
\begin{equation}
N=\int_0^{r_0} 4 \pi n_e r^2 dr= Z \int_0^{x_0} \frac{\chi^{3/2}}{x^{1/2}} \left[1+ \left(\frac{Z}{Z_{cr}}\right)^{4/3}\frac{\chi}{x} \right]^{3/2}\;\; x \;\; dx
\end{equation}
with $r_0=bx_0$ atom size. Developing this formula we have
\begin{equation}
N = Z \int_0^{x_{nuc}} x \chi'' \;\; dx + \frac{3Z}{x_{nuc}^3} \int_0^{x_{nuc}} x^2 \;\; dx + Z \int_{x_{nuc}}^{x_0} x \chi'' \;\; dx
\end{equation}
which gives the simple relation
\begin{equation}
N=Z\left[x_0\chi'(x_0)-\chi(x_0)+1 \right]
\label{con2}
\end{equation}

For a neutral atom we have $N=Z$ and the condition \ref{con2} reads simply
\begin{equation}
x_0\chi'(x_0)=\chi(x_0)
\label{con3}
\end{equation}

Note that the physical quantities of interest, such as the coulomb potential and the density of electrons do not show any singularity in the center, neither on the border of the nucleus, being dependent just on the function $\chi$ and his first derivative. At the opposite the only discontinuity appears in the second derivative of $\chi$ due to our rough assumption of homogeneous spherical nucleus.

\begin{figure}[t]
\resizebox{\hsize}{12cm}{\includegraphics{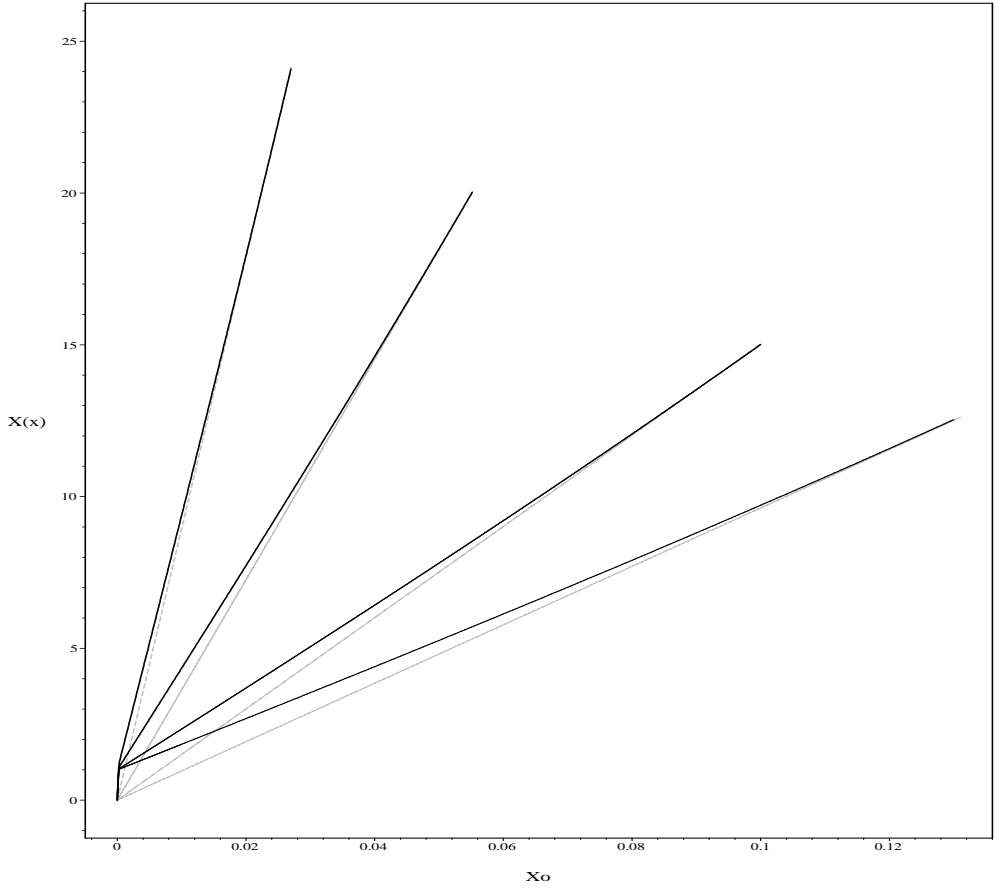}}
\caption[fig]{Solution of the generalised Fermi-Thomas adimensional equation for a Fe atom in several states of compression.}
\label{fig.thofe}
\end{figure}

We show in figure \ref{fig.thofe} a numerical integration of \ref{thomfer} for a  $^{56}_{26}Fe$ atom, in several states of compression.

\begin{figure}

\begin{center}

\begin{narrowtabular}{3cm}{rrll}
\hline\hline
$P_{Fe}$(dyne/$cm^2$)    &  $\rho_{Fe}$(g/$cm^3$)  &   $P_{He}$(dyne/$cm^2$)    &  $\rho_{He}$(g/$cm^3$)   \\
\hline 

& \\
$4,577\cdot 10^{22}$& 	$1,624\cdot 10^{6}$ &$1.771\cdot 10^{20}$	&$2.537\cdot 10^{1}$\\

$9,055\cdot 10^{22}$& 	$2,523\cdot 10^{6}$ &$4.077\cdot 10^{20}$	&$4.199\cdot 10^{1}$\\

$3,061\cdot 10^{23}$& 	$5,640\cdot 10^{6}$ &$7.621\cdot 10^{20}$	&$6.140\cdot 10^{1}$\\

$4,053\cdot 10^{24}$& 	$3,368\cdot 10^{7}$ &$2.745\cdot 10^{21}$	&$1.346\cdot 10^{2}$\\

$7,447\cdot 10^{25}$& 	$2,787\cdot 10^{8}$ &$7.938\cdot 10^{22}$	&$1.129\cdot 10^{3}$\\

$2,132\cdot 10^{26}$& 	$6,067\cdot 10^{8}$ &$3.537\cdot 10^{23}$	&$3.045\cdot 10^{3}$\\

$6,193\cdot 10^{26}$& 	$1,341\cdot 10^{9}$ &$9.613\cdot 10^{23}$	&$6.030\cdot 10^{3}$\\

$1,179\cdot 10^{27}$& 	$2,168\cdot 10^{9}$ &$2.060\cdot 10^{24}$	&$1.026\cdot 10^{4}$\\

$2,854\cdot 10^{27}$& 	$4,195\cdot 10^{9}$ &$7.313\cdot 10^{24}$	&$2.526\cdot 10^{4}$\\

$5,875\cdot 10^{27}$& 	$7,198\cdot 10^{9}$ &$3.092\cdot 10^{25}$	&$7.193\cdot 10^{4}$\\

$1,837\cdot 10^{28}$& 	$1,690\cdot 10^{10}$ &$1.166\cdot 10^{26}$	&$1.911\cdot 10^{5}$\\

$2,930\cdot 10^{28}$& 	$2,398\cdot 10^{10}$ &$3.118\cdot 10^{26}$	&$3.964\cdot 10^{5}$\\

$4,451\cdot 10^{28}$& 	$3,280\cdot 10^{10}$ &$6.834\cdot 10^{26}$	&$7.110\cdot 10^{5}$\\

$9,180\cdot 10^{28}$& 	$5,641\cdot 10^{10}$ &$1.314\cdot 10^{27}$	&$1.158\cdot 10^{6}$\\

$1,694\cdot 10^{29}$& 	$8,928\cdot 10^{10}$ &$2.302\cdot 10^{27}$	&$1.760\cdot 10^{6}$\\

$2,880\cdot 10^{29}$& 	$1,329\cdot 10^{11}$ &$6.474\cdot 10^{28}$	&$2.141\cdot 10^{7}$\\

$6,998\cdot 10^{29}$& 	$2,586\cdot 10^{11}$ &$3.882\cdot 10^{29}$	&$8.197\cdot 10^{7}$\\

$1,989\cdot 10^{30}$& 	$5,658\cdot 10^{11}$ &$1.331\cdot 10^{30}$	&$2.065\cdot 10^{8}$\\

$3,518\cdot 10^{30}$& 	$8,674\cdot 10^{11}$ &$7.299\cdot 10^{30}$	&$7.397\cdot 10^{8}$\\

$1,107\cdot 10^{31}$& 	$2,048\cdot 10^{12}$ &$2.399\cdot 10^{31}$	&$1.805\cdot 10^{9}$\\

$2,696\cdot 10^{31}$& 	$3,988\cdot 10^{12}$ &$5.995\cdot 10^{31}$	&$3.588\cdot 10^{9}$\\

$4,237\cdot 10^{31}$& 	$5,594\cdot 10^{12}$ &$2.365\cdot 10^{32}$	&$1.004\cdot 10^{10}$\\

$1,032\cdot 10^{32}$& 	$1,089\cdot 10^{13}$ &$1.004\cdot 10^{33}$	&$2.966\cdot 10^{10}$\\

$9,468\cdot 10^{32}$& 	$5,693\cdot 10^{13}$ &$2.099\cdot 10^{33}$	&$5.150\cdot 10^{10}$\\

$2,159\cdot 10^{33}$& 	$1,050\cdot 10^{14}$ &$4.503\cdot 10^{33}$	&$9.118\cdot 10^{10}$\\

$5,546\cdot 10^{33}$& 	$2,109\cdot 10^{14}$ &$5.163\cdot 10^{33}$	&$1.010\cdot 10^{11}$ \\

& \\
\hline 
\hline  
\end{narrowtabular}
\vskip 1cm
\caption{Equation of state obtained for Iron (the two columns on the right) and for Helium. Pressure and mass densities can be obtained directly the quantities in tab.1 (see text).}
\end{center}
\label{tabella}
\end{figure}

\section{The equation of state}

The study of the previous section allows us to determine, under suitable hypothesis, the equation of state of compressed matter, by simply computing for each value of the atomic compression parameter $x_0$ the corresponding  values of pressure and mass density. The origin of these two physical quantities is evidently different, the first being generated from the degenerate gas of electrons and the second being essentially given by the mass density of nuclei. 
We shall assume that nuclei are arranged in a Wigner-Seitz lattice, each cell being filled by a relativistic gas of degenerate electrons.  As shown by Salpeter \cite{salpeter} the shape of  the lattice cell is in first approximation unimportant, so we'll assume a spherical cell, with a radius equal to the atomic radius.  The  average density of cells is thus equal to
\begin{equation}
\overline{\rho}(x_0)= \frac{Am_0}{V_{cell}} = \frac{3 Am_0}{4 \pi (b x_0)^3}
\end{equation}
where A and $m_0$ are respectively the number and the mass of nucleons.

In this scheme also the pressure is easily determined, being simply the pressure generated by a Fermi gas of number density equal to the number density of electrons at the border of the atom $n_e(x_0)$, which can be expressed in terms of adimensional quantities as follows

\begin{equation}
n_e(x_0)=\frac{Z}{4\pi b^3}\left(\frac{\chi(x_0)}{x_0}\right)^{3/2}\left[1+ \left(\frac{Z}{Z_{cr}}\right)^{4/3}\frac{\chi(x_0)}{x_0} \right]^{3/2} 
\end{equation}

In tab.\ref{tabella} we list the resulting equation of state for different atomic species.

\begin{figure}[t]
\resizebox{\hsize}{!}{\includegraphics{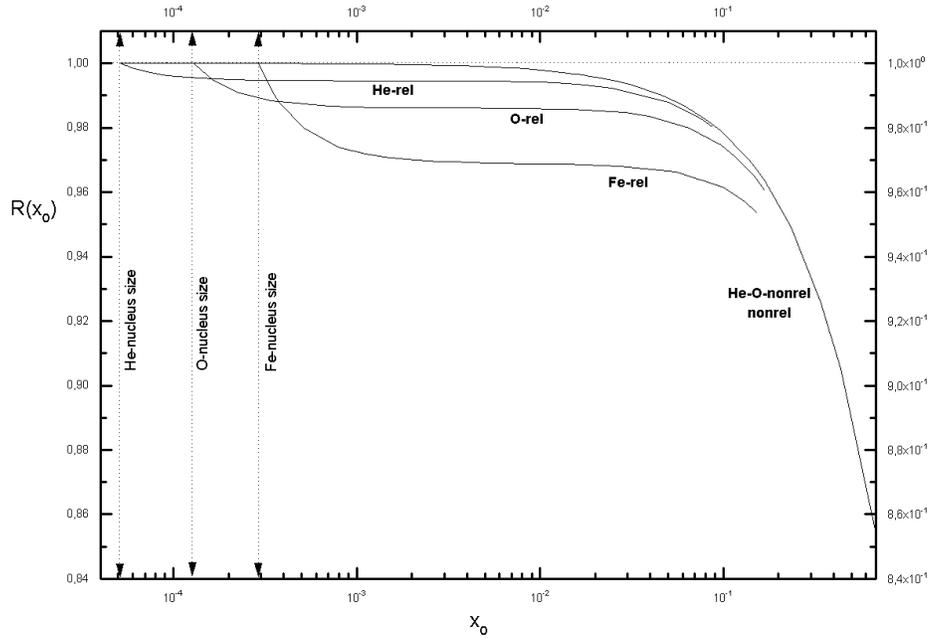}}
\caption[fig]{Ratio $R(x_0)$ between density at the border and average density for Fe, O and He atoms. It is shown also the corresponding results obtained with a non-relativistic treatment.}
\label{dialbordo}
\end{figure}

We would like to stress the deep difference between our model and the Chandrasekhar's one, evidently corresponding to an electron gas uniformly distributed within the cell, by showing the  behaviour (see figure \ref{dialbordo}) of the ratio $R(x_0)$ between the density at the border and the average number density, given by

\begin{equation}
R \equiv \frac{n_e(x_0)}{\overline{n}(x_0)}=\frac{1}{3} \left( x_0 \chi(x_0) \right)^{3/2} \left[ 1+  \left(\frac{Z}{Z_{cr}} \right) ^{4/3} \frac{\chi(x_0)}{x_0} \right] ^{3/2}
\end{equation}
The analysis of this plot clarifies some important features: 
\begin{itemize}
\item For high values of $x_0$, i.e. for low densities the ratio $R$ reduces to the non-relativistic (dashed line) expression 

\begin{equation}
R_{nonrel}(x_0) =\frac{1}{3} \left( x_0 \chi(x_0) \right)^{3/2} 
\end{equation}

\item For lower value of $x_0$,  corresponding to values of density  up to an important fraction of nuclear density one gets the ultra-relativistic value

\begin{equation}
R_{ultrarel} =\frac{1}{3} \left(\frac{Z}{Z_{cr}} \right) ^{2} \chi(x_0)^3
\end{equation}

It can be shown that scaling laws apply in this case and that the ratio is in effect independent from the compression parameter (see \cite{preparation}).

\item Finally for $x_0=x_{nuc}$ we recover the uniformity of the electron distribution.
\end{itemize}

But our treatment allows us not only to remove the zero-order approximation of Chandrasekhar (uniformly distributed electrons) but also the first-order approximation of Salpeter who could not solve the TF equation for each value of Z and $x_0$, as is done here, and introduced a development of the number density of electrons
\begin{equation}
n(x)= n_0 + \epsilon(x)
\end{equation}
which is evidently only allowed when the electrostatic interaction within nuclei and electrons is small respect to the Fermi Energy of electrons, i.e. in the limit of high densities.
 
\section{Equilibrium configurations of white dwarfs} 

We are now able to compute equilibrium configurations of white dwarfs. It is clear that we will obtain, starting with the same central density, less massive configurations than those obtained by both Chandrasekhar and Salpeter. 
That is due to the fact that for a fixed mass density, i.e. for a fixed value of A and of the compression parameter, our pressure is systematically lower being generated by a shell of electrons whose density is, as we shown, always lower than the average one and eventually very far from the mean value.

Numerical integration of the models are here performed for stars of Helium and Iron in the framework of a Newtonian theory of gravitation and General Relativity. In the last case we have the usual TOV equation of hydrostatic equilibrium
\begin{eqnletter}
\frac{d M(r)}{dr}= 4 \pi r^2 \epsilon(r)   \\
\frac{dp}{dr} = - \frac{(p+ \epsilon)}{c^2}\frac{ \left( \frac{G M(r)}{r^2} + \frac{4 \pi G}{c^2} r\; p \right)}{1- \frac{2 G M(r)}{c^2 r}}            
\label{tre}
\end{eqnletter}
and we add to these equations the numerical equation of state obtained in the previous section. 

\begin{figure}[t]
\resizebox{\hsize}{!}{\includegraphics{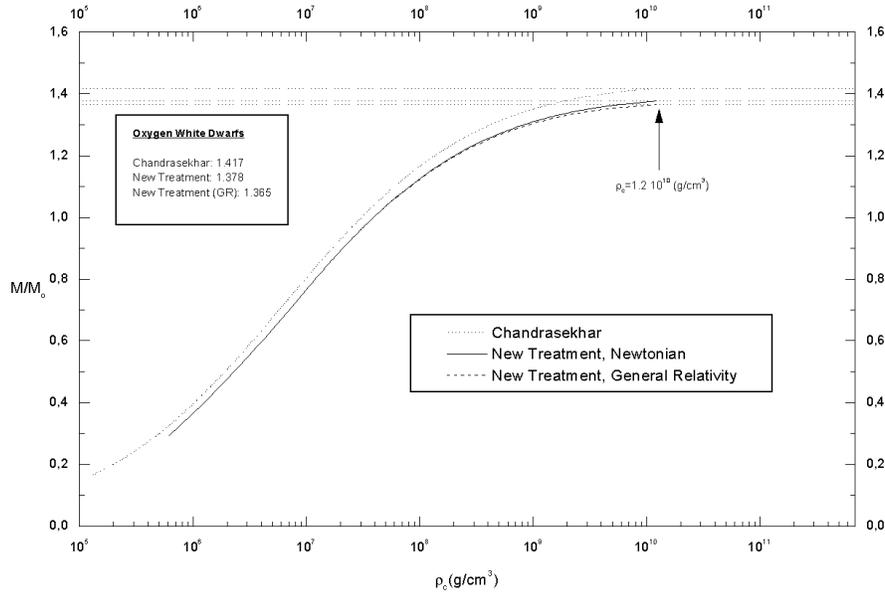}}
\caption[fig]{Equilibrium configurations curve for Oxygen WD in the $M/M_{\odot} - \rho_c$ plane, obtained in Newtonian theory and General relativity. For comparison we show the results obtained with Chandrasekhar model.}
\label{ossigeno}
\end{figure}

\begin{figure}[t]
\resizebox{\hsize}{!}{\includegraphics{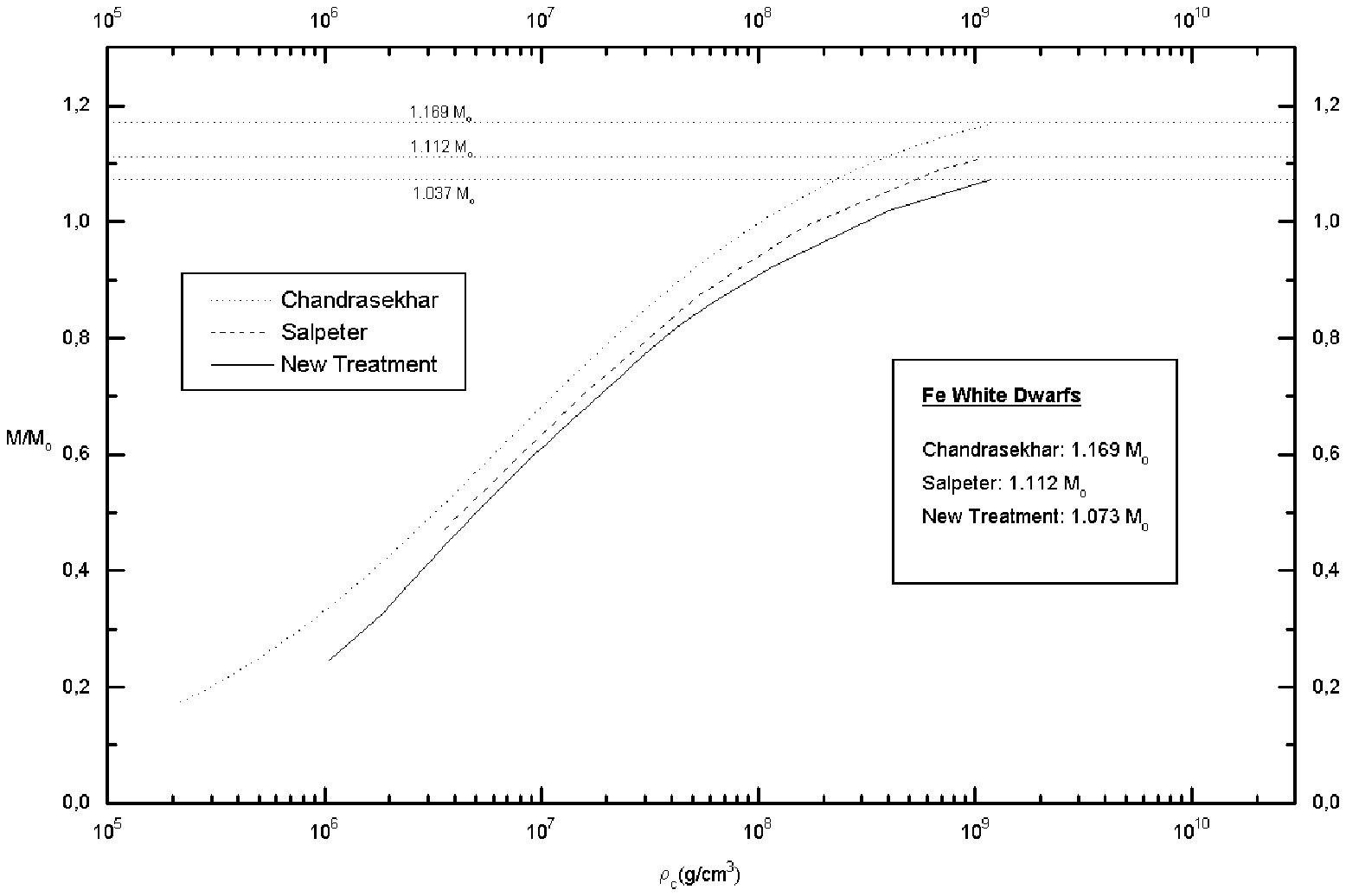}}
\caption[fig]{Equilibrium configurations curve for Iron WD in the $M/M_{\odot} -  \rho_c$ plane, obtained in Newtonian theory (the General Relativistic curve is practically superposed on the newtonian one at these low densities). For comparison we show the results obtained by Chandrasekhar and Salpeter.}
\label{ferro}
\end{figure}

Results of integrations, plotted in figg.\ref{ossigeno} and \ref{ferro}, show that deviations from classical models become very important at the lower densities. 

The mass for an Iron star  corresponding to a central density of the order of $10^6 g/cm^3$ is about 30\% smaller than what predicted from the Chandrasekhar model and about 10\% smaller than the Salpeter one. Of course deviations are most marked for atomic specie with high Z, being related, as seen, to Coulomb corrections to the equation of state.
Numerical integration of equilibrium configurations can be computed up to arbitrarily high central densities, but to make the model physically reasonable we have to make allowance to the occurrence of nuclear reactions. In particular when the Fermi Energy of electrons is sufficiently high they can undergo inverse beta decay, consisting in a reaction of the type
\[ e^- + p \rightarrow n + \nu_e \]
Using the recent (1993) nuclear data in Audi et al \cite{audi} we can find for each element the beta decays energies, and thus the critical density at which the element considered undergoes inverse beta decay. As first shown in Harrison et al \cite{wheeler}, the introduction of nuclear reactions leads to the existence of a maximum mass which separate stable from unstable configurations (see \cite{wheel2}, based on the fundamental works of Chandrasekhar \cite{ch64a}\cite{ch64b}). 
Using this data, we stop the curve when the central density is equal to the critical density. The configuration wich correspond to the last point is the maximum mass configuration.  For an Oxygen and an Iron WD, we find respectively $1.365   M_{\odot}$ and $1.063   M_{\odot}$. These values differ roughly about 5\% from the Salpeter configurations and 10\% from those evaluated by Chandasekhar (see figures).

\section{Conclusions}
The derivation of the relativistic generalised Fermi-Thomas model is here presented, and a FMT treatment of the compressed atom is applied. Numerical examples are given of compressed atomic configurations and an equation of state is derived for a zero temperature plasma, removing some unneeded approximations applied in classical papers. Results are applied to the problem of equilibrium configurations of White Dwarfs, described as cold degenerate stars and a class of configurations is obtained (for the same central densities)  less massive  than those predicted in the models of Chandrasekhar and Salpeter; in particular we give an estimate of the maximum mass for Oxygen and Iron WD.

\end{document}